\documentclass[aps,prb,twocolumn,groupedaddress]{revtex4-1}

\usepackage{float}
\usepackage{graphicx}
\usepackage{caption}
\usepackage{subcaption}
\usepackage[export]{adjustbox}

\newcommand{\figref}[1]{Fig.~\ref{#1}}

\newcommand{\ket}[1]{\ensuremath{|{#1}\rangle}}

\begin{document}

\title{Nitrogen vacancy centers in diamond as angle-squared sensors}


\author{Shonali Dhingra}
\author{Brian D'Urso}
\email[]{dursobr@pitt.edu}
\affiliation{Department of Physics and Astronomy, University of Pittsburgh, 3941 O'Hara Street, Pittsburgh, PA-15260, USA }


\date{\today}

\begin{abstract}
Nitrogen-vacancy (NV) centers are defects in diamonds, which, due to their electronic structure, have been extensively studied as magnetic field sensors. Such field detection applications usually employ the NV centers to detect field components aligned with the direction of the internally-defined spin axis of the NV center. In this work we detect magnetic fields which are slightly misaligned with the NV center axis. In particular, we demonstrate that the NV center can measure the square of the angle between the magnetic field and the NV center axis with high sensitivity which diverges as the external field approaches a value pre-defined by NV center's internal parameters, in agreement with predictions. These results show that NV centers could be used as sensitive transducers for making quantum nondemolition (QND) measurements on systems such as nanomechanical oscillators.
\end{abstract}


\maketitle

\section{\label{sec:Introduction}Introduction}

A variety of hybrid systems involving mechanical oscillators coupled with spin systems have recently been studied and used for characterization of quantum properties of their components~\cite{Rugar04, Arcizet2011, Durso11}. With the advent of such systems, a class of measurement known as quantum nondemolition (QND) is re-emerging. Since the introduction of QND measurements~\cite{Braginsky80,Caves80}, this measurement class has found use in a wide variety of applications such as gravitational-wave detection~\cite{Brillet1985}, weak force detection using harmonic oscillators~\cite{Bocko96}, optics~\cite{Grangier1998}, etc. QND measurements allow repeated quantum measurements of certain quantities separated in time to give easily understood results. In a previous work, we proposed a hybrid system comprising of a nanomechanical oscillator (NMO) coupled to a nitrogen vacancy (NV) center in presence of an external magnetic field with a QND interaction between the NV center and NMO. The QND measurement results from the quadratic coupling between the NMO and the NV center, which was predicted to have high sensitivity~\cite{Durso11}.

NV centers are naturally occurring defects in diamond, which consist of a substitutional nitrogen atom and an adjacent carbon vacancy~\cite{Loubser1978}. These defects can also be artificially introduced in diamonds through irridiation and annealing~\cite{Manson06, Doherty2013}. The as-formed neutral NV center acquires an additional electron from elsewhere in the lattice, thus having six electrons~\cite{Manson06}. This energetically favorable charged state of the NV (NV$^{-}$) will be referred to as NV throughout this work. Owing to their ease of manipulation and readout, and long coherence times, they have recently being extensively characterized and used in various applications~\cite{Doherty2013}. One such application of NV centers is as a sensor for steady and oscillating magnetic fields~\cite{Degen08, Taylor08, Balasub08, Maze08, Steinert2010, Hall2009}. These previous works have shown that the NV centers can be used to detect nanoTesla magnetic fields with nanoscale spatial resolution at temperatures below 4~K~\cite{Maze08} or at room temperature~\cite{Balasub08}, and can be used to map out the magnetic fields in a two-dimensonal area~\cite{Steinert2010}. Most of these applications choose the NV center aligned in the direction of the external magnetic field~\cite{Maze08} or they detect the component of the magnetic field that is aligned in the direction of the NV center~\cite{Taylor08}. Applications that detect magnetic fields that are misaligned with the direction of the NV center do so by iteratively fitting parameters and minimizing error functions~\cite{Steinert2010}. Due to these limitations, the maximum magnetic field measured by such techniques has been limited to $\leq$10~mT~\cite{Balasub08, Steinert2010}.

In this work, we experimentally demonstrate that an NV center in an external magnetic field can be used as a detector with quadratic coupling to the angle $\theta$ between the magnetic field and the NV center axis. Instead of simply considering the NV center to be a sensor for the magnetic field in a single direction, in this work we experimentally confirm a fuller but simple quantum treatment of the coupling which predicts that an NV center can be used to measure $\theta^{2}$ with sensitivity which diverges as the magnetic field approaches a critical value, which may allow the NV center to be used for QND measurements in nanomechanical systems. We briefly summarize the previously reported theoretical results~\cite{Durso11} below.



\section{\label{sec:Theory}Theory}

\begin{figure}
\centering
	\includegraphics[scale=0.17]{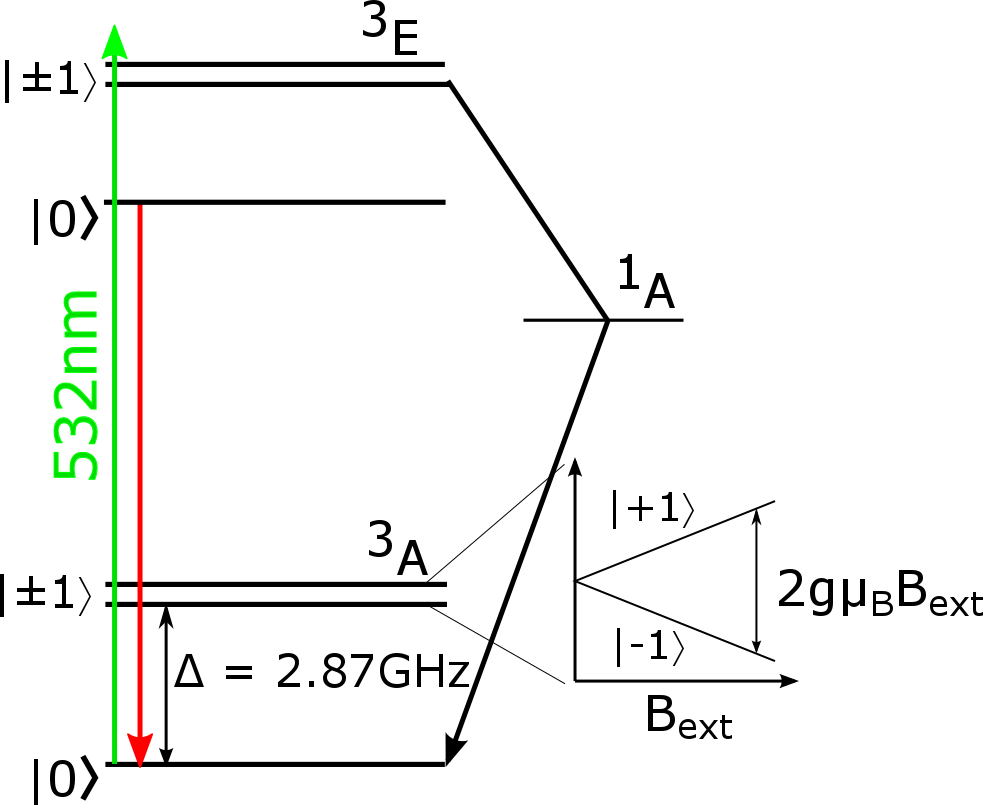}
\caption{Energy structure of the ground state of the NV center.}
\label{fig:NVCenter}
\end{figure}

The energy structure of the ground state of an NV center has been extensively studied~\cite{Loubser1978, Doherty2013}, and is understood to be as shown in~\figref{fig:NVCenter}. The ground state ($^{3}A$) is a spin triplet, with $\ket{0}$, $\ket{1}$ and $\ket{-1}$ spin sublevels, defined in the $S_z$ basis ($\hat{z}$ being along the NV center axis), where state $\ket{i}$ has $m_s$ = $i$. 
In the absence of any external magnetic field, $\ket{1}$ and $\ket{-1}$ are degenerate, with a zero-field splitting of $\Delta=$~2.87~GHz above the $\ket{0}$ state. In presence of an external magnetic field, the degeneracy between the $\ket{1}$ and $\ket{-1}$ states is lifted by Zeeman splitting. 

The Hamiltonian for the ground state of the NV center can be written as~\cite{Wrachtrup06rev, Taylor08}:
\begin{equation}
H_\mathit{NV}=\hbar \Delta S_z^2 + g\mu_B \left(\vec{S}\cdot\vec{B}_\mathit{ext}\right)
\label{eq:hnv0}
\end{equation}
where $\Delta$ is the zero-field splitting, $g\approx 2$ is the electron g-factor in the NV center, $\mu_B$ is the Bohr magneton, $\vec{S}$ is the electronic spin of the NV center and $\vec{B}_\mathit{ext}$ is the external magnetic field. 

In our experiment, the applied magnetic field $\vec{B}_\mathit{ext}$ may not be aligned with the NV spin vector $\vec{S}$. This gives the term $\vec{S}\cdot\vec{B}_\mathit{ext}$ in Eq. \ref{eq:hnv0} an angular ($\theta$) dependence, where $\theta$ is the angle between $\vec{B}_\mathit{ext}$ and $\vec{S}$. For this work, we fix the direction of the NV center, and vary the external magnetic field. For the NMO-NV proposed hybrid system~\cite{Durso11}, the direction of magnetic field is fixed and the direction of the NV center moves due to the oscillation of the NMO. In either case, $H_{NV}$ takes the following form, where we assume $\vec{B}_\mathit{ext}$ is in the $y-z$ plane:
\begin{equation}
H_\mathit{NV}=\hbar \Delta S_z^2 + g\mu_B B_\mathit{ext} \left(S_y \sin\theta + S_z \cos\theta \right)
\label{eq:hnv}
\end{equation}

%
%

The eigenvalues of $H_\mathit{NV}$,  $\lambda_i (\theta)$, with $\theta$ as a parameter, have the form
\begin{equation}
\label{eq:EigEnergies}
\lambda_i (\theta) = \hbar\omega_i + \kappa_i\theta^2 + O(\theta^4)
\end{equation}
where $i=$ $-1$, $0$, or $+1$ is a label which is equivalent to the NV center $m_s$ when $\theta=0$.
$\hbar\omega_i$ is simply the result of combining the zero field splitting with the static external field at $\theta=0$:
\begin{eqnarray}
\label{eq:omega_-1}
\omega_{\pm 1}&=& \Delta \left(1 \pm B_\mathit{ext}/B_\mathit{zfs} \right) \\
\omega_0 &=& 0
\end{eqnarray}
where $B_\mathit{zfs}=\hbar\Delta/g\mu_B \approx 102.5$~mT is the magnitude of the effective internal magnetic field which results in the zero field splitting $\Delta$. 

The $\theta^2$ coefficient in $\lambda_i (\theta)$, $\kappa_i$,
is:
\begin{eqnarray}
\kappa_{\pm 1} &=& - \hbar\Delta \frac{B_\mathit{ext}}{2(B_\mathit{ext}\pm B_\mathit{zfs})} \\
\kappa_0 &=& \hbar\Delta \frac{B_\mathit{ext}^2}{B_\mathit{ext}^2-B_\mathit{zfs}^2}
\end{eqnarray}
Therefore, the $\theta^2$ dependence of the energy gap between $\ket{0}$ and $\ket{-1}$, and $\ket{0}$ and $\ket{+1}$  near $\theta = 0$ ($\kappa_0-$ and $\kappa_0+$, respectively) are
\begin{eqnarray}
\kappa_{0-} &= - \hbar\Delta \frac{(B_\mathit{ext})(3B_\mathit{ext} + B_\mathit{zfs})}{2(B_\mathit{zfs}^2-B_\mathit{ext}^2)} \\
\kappa_{0+} &= - \hbar\Delta \frac{(B_\mathit{ext})(3B_\mathit{ext} - B_\mathit{zfs})}{2(B_\mathit{zfs}^2-B_\mathit{ext}^2)}
\end{eqnarray}
The denominators of $\kappa_0-$ and $\kappa_0+$ suggest that the sensitivity of the spin transition frequencies to $\theta^2$ around $\theta=0$ should increase much faster than linearly and even diverge as $B_\mathit{ext}$ approaches $B_\mathit{zfs}$, the result of an avoided level crossing in the NV center.

\section{\label{sec:Expt}Experimental Data}

To test the predictions, electronic spin resonance (ESR) was performed on diamond nanocrystals with NV center ensembles. The nanocrystals used for this purpose were $\sim$100~nm and  were prepared using high pressure high temperature (HPHT) techniques (Ad\'{a}mas Nanotechnologies: ND-NV-100nm (400NV)). The diamond nanocrystals were spin-coated on polished silicon with low density, so individual nanocrystals could be resolved. Confocal laser scanning microscopy (CLSM) was employed to visualize and detect the photoluminescence of the NV centers. Green laser light ($\lambda$ = 532~nm) was used to excite the NV centers in these nanocrystals and their photoluminescence was detected using a single photon counting module (SPCM) after filtering out wavelengths between 517~nm and 548~nm using a notch filter. To drive resonant transitions between $\ket{0}$ and $\ket{\pm 1}$ spin states, a microwave (MW) drive was applied to this sample using an un-terminated loop, 2 to 3~mm in diameter, made out of 30~$\mu$m diameter gold-plated tungsten wire. A dip in the photoluminescence intensity is detectable when the NV centers are driven at a frequency resonant with a spin transition. In the absence of an external magnetic field a characteristic ESR signal, a dip in photoluminescence at $\sim$2.87~GHz is expected from single NV centers or NV center ensembles~\cite{Gruber97, Balasub08, Doherty2013}, and was obtained on our samples as well. 



An external magnetic field $B_\mathit{ext}$ with continuously variable magnitude and direction was applied to the nanodiamonds using three pairs of electromagnetic coils. Due to experimental limitations, $\approx \pm$100~mT was available in the direction perpendicular to the sample ($\hat{z}^\prime$), and $\approx \pm$25~mT in each of the two directions parallel to the sample surface ($\hat{x}^\prime$ and $\hat{y}^\prime$). The range of accessible magnetic fields is illustrated graphically in Fig. 2(a). For weak fields, $B_\mathit{ext} \leq$ 25~mT, the external magnetic field could be applied in any direction, but for $B_\mathit{ext}$ approaching $B_\mathit{zfs}$, only a small angular range was accessible. 


\begin{figure*}
\centering
\includegraphics[scale=0.33]{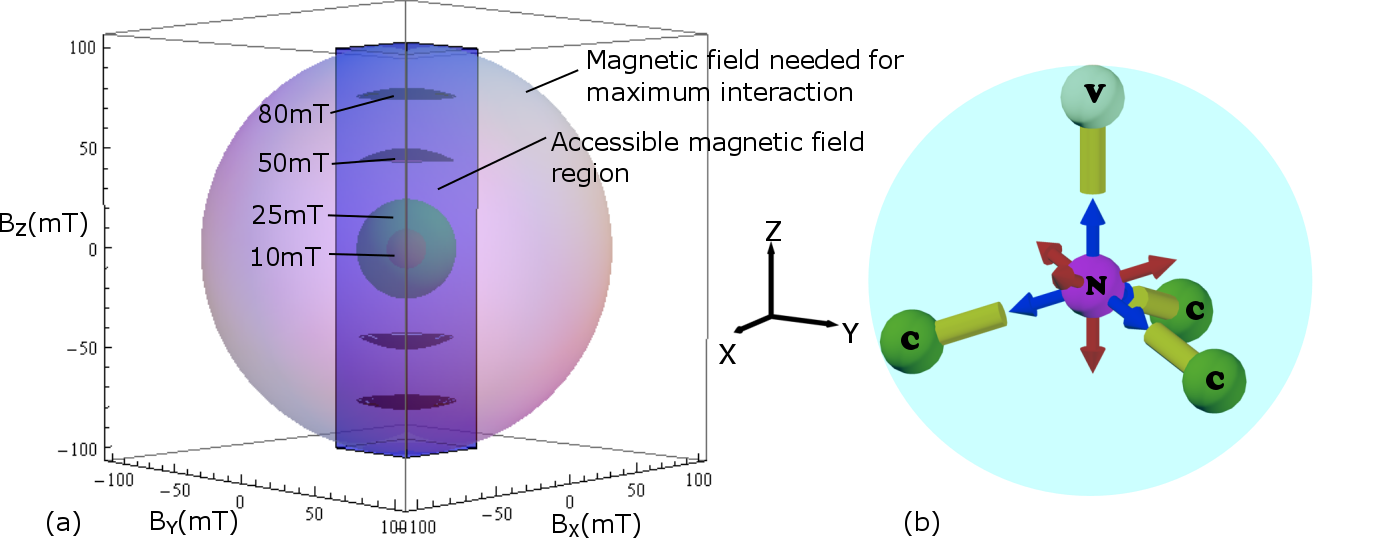}
\caption{Pictorial 3-D representation of experimentally accessible external magnetic field ($B_\mathit{ext}$) and possible NV center orientations (directions of $B_\mathit{zfs}$) in any arbitrary diamond nanocrystal (a) The pink sphere represents the direction of possible $B_\mathit{zfs}$ for all diamond nanocrystals. The cuboid with square faces of side 25~mT and a height of $\pm$100~mT represents the experimentally accessible magnetic field region, along the experimentally defined $\hat{z}^\prime$ direction of $B_\mathit{ext}$. This limited the angular range of a certain magnitude of magnetic field we could apply around the sample, as shown by the shapes inside the cuboid. (b) A carbon tetrahedron structure in any arbitrary diamond nanocrystal is shown. Assuming the nitrogen atom is at the center of the tetrahedron (pink), and one of the other carbon atom (green) sites is vacant (light green), the NV-axis could be directed along one of four possible orientations (shown in blue arrows). Due to the symmetry between $\ket{1}$ and $\ket{-1}$ spin states, the NV-axis could also be directed along the four orientations opposing the carbon tetrahedron bonds (shown in red arrows). 
}
\label{fig:3D}
\end{figure*} 

Individual diamond nanocrystals were randomly oriented on the sample. Within each nanocrystal, the direction of an NV center axis can be in any one of the four possible orientations along the carbon tetrahedron bonds, defining the NV center $\hat{z}$ direction (shown with blue arrows in Fig. 2(b)). Due to the symmetry between $\ket{1}$ and $\ket{-1}$ states, the NV $\hat{z}$ direction could also be chosen to be any of the four orientations directly opposite to the carbon tetrahedron bonds (shown in red arrows in in Fig. 2(b)). If each diamond nanocrystal has many NV centers with random orientations, there are eight possible directions where $B_\mathit{zfs}$ and $B_\mathit{ext}$ could align with each other, each selecting a subset of the NV centers. These eight directions define the orientations along which we expect the NV centers could behave as $\theta^{2}$ sensors with very high sensitivity. 

\begin{figure*}
	\begin{subfigure}{0.32\linewidth}
		\includegraphics[scale=0.2]{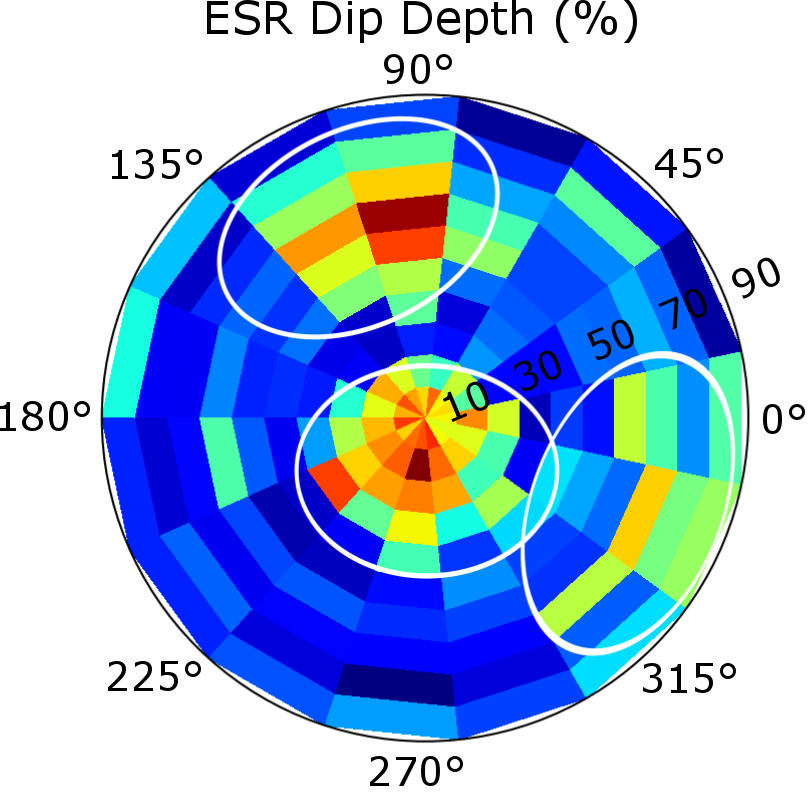}
		\caption{\label{fig:20mTDepth}}
	\end{subfigure}
	\begin{subfigure}{0.32\linewidth}
		\includegraphics[scale=0.2]{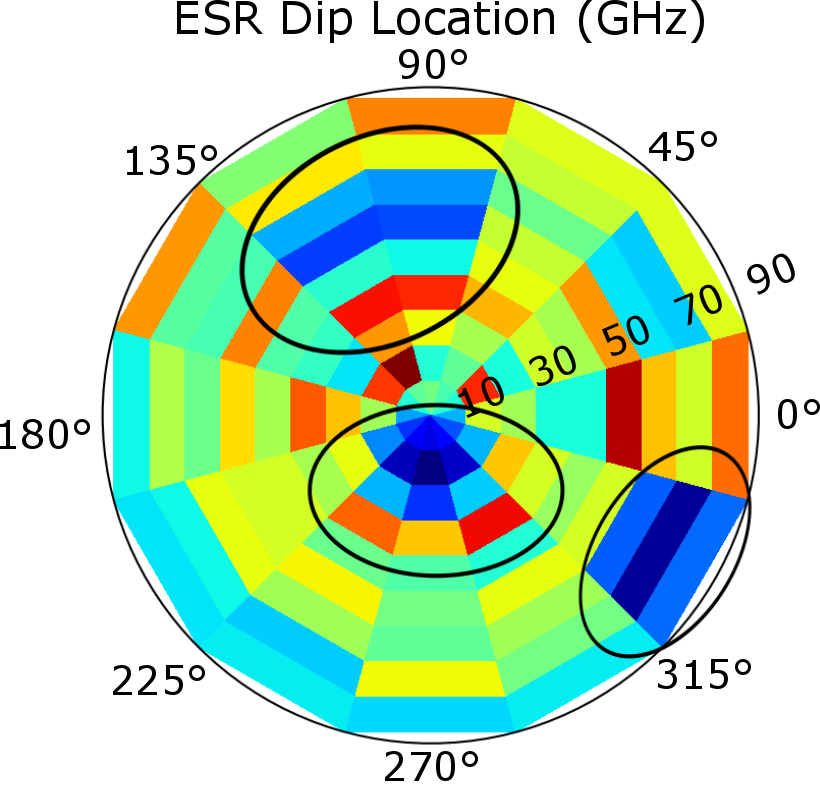}
		\caption{\label{fig:20mTLocation}}
	\end{subfigure}
	\begin{subfigure}{0.32\linewidth}
		\includegraphics[scale=0.15]{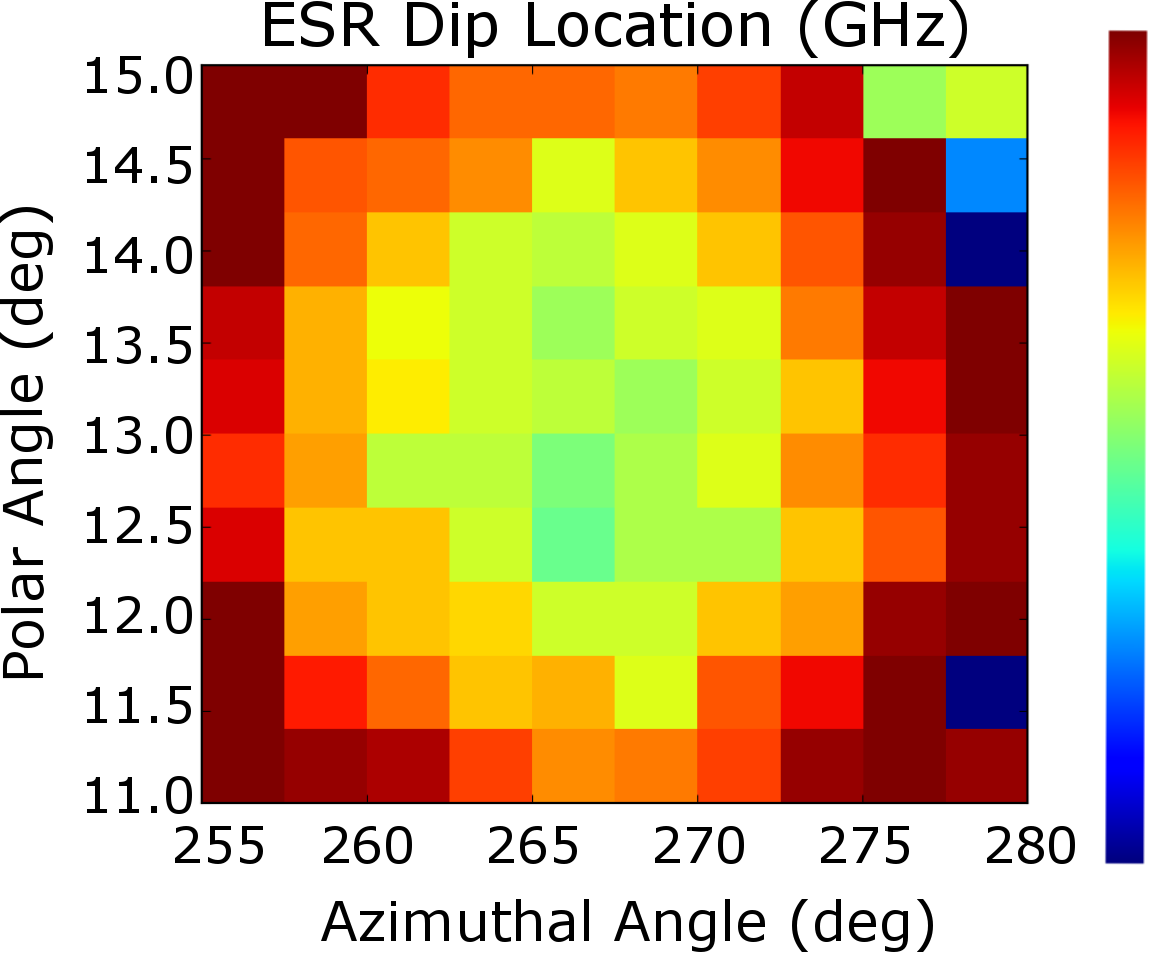}
		\caption{\label{fig:80mTLocation}}
	\end{subfigure}
	\\
	\begin{subfigure}{0.495\linewidth}
		\includegraphics[scale=0.25]{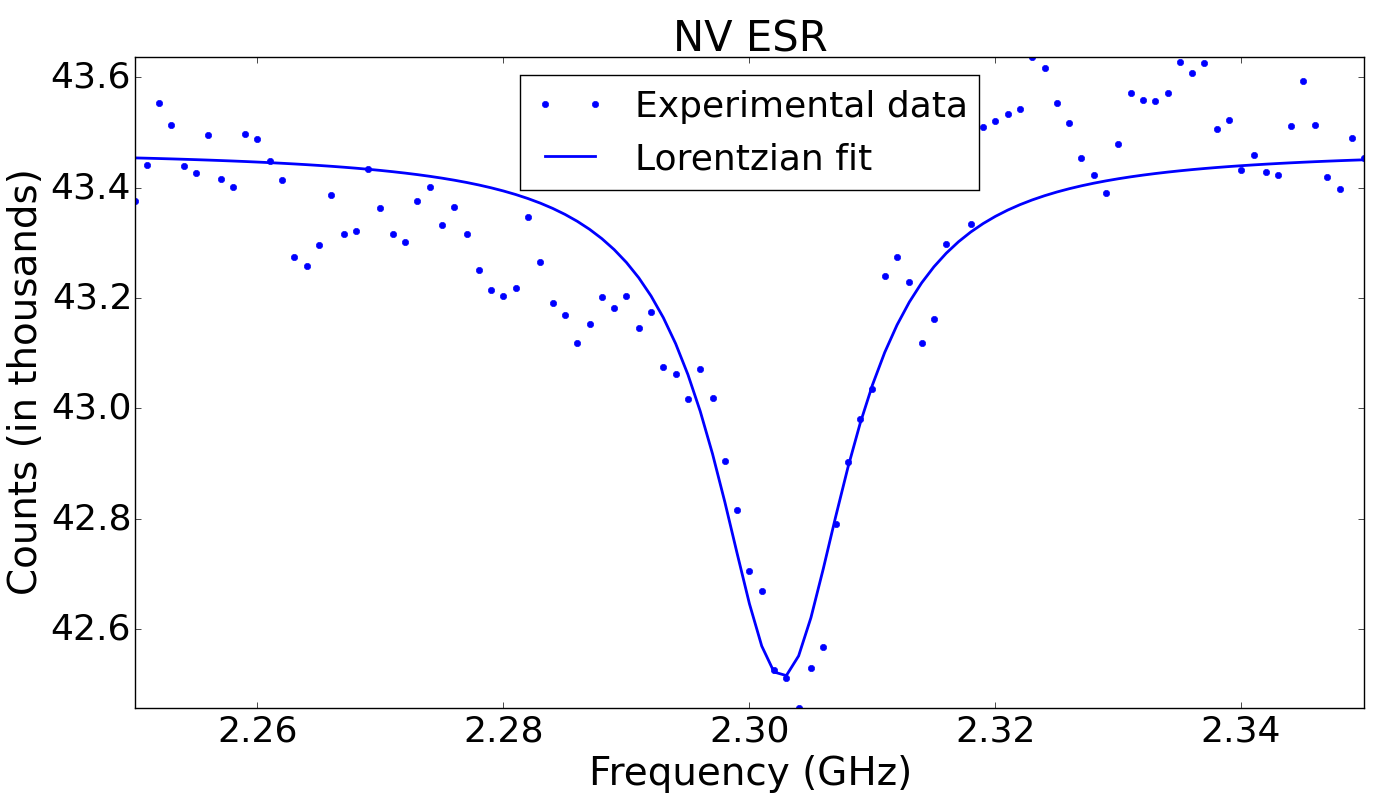}
		\caption{\label{fig:20mTDip}}
	\end{subfigure}
	\begin{subfigure}{0.495\linewidth}
		\includegraphics[scale=0.25]{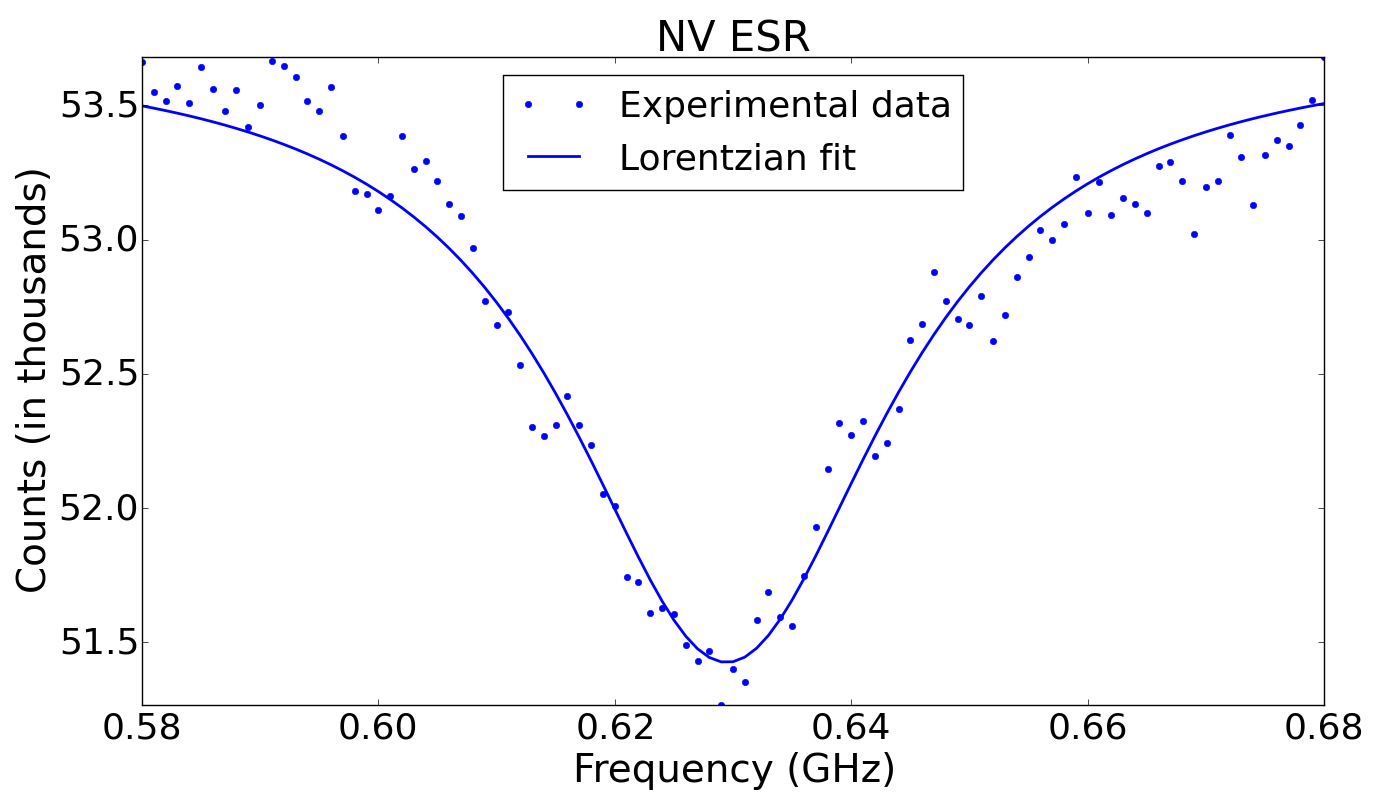}
		\caption{\label{fig:80mTDip}}
	\end{subfigure}
	
\caption{Locating NV orientations for an arbitrarily chosen diamond nanocrystal: (\subref{fig:20mTDepth}) monitoring the ESR dip depth while sweeping 20~mT $B_\mathit{ext}$ around in azimuthal and polar angles about the $+\hat{z}$ direction and projecting the hemisphere onto a polar plot and (\subref{fig:20mTLocation}) similarly monitoring and plotting the ESR dip location. Circled regions in (\subref{fig:20mTDepth}) and (\subref{fig:20mTLocation}) exhibit maximum dip depth and minimum dip frequency, which can be thought of as orientations of the NV centers in a diamond nanocrystal. (\subref{fig:80mTLocation}) Similarly monitoring the ESR dip location at 80~mT $B_\mathit{ext}$ and projecting the spherical cap onto a planar plot. The colormap for the projection plots is also shown.~(\subref{fig:20mTDip}) ESR dip at 20~mT in a region closest to the pole, where the dip location reaches a minimum in frequency (12.5$^\circ$~$\phi$ and 264$^\circ$~$\Theta$). (\subref{fig:80mTDip}) ESR dip at 80~mT at the same location as above, where the dip frequency reaches a minimum.}
\label{fig:BField}
\end{figure*}

Experimentally, the first task was to identify the NV axis orientations in a particular diamond nanocrystal by finding the magnetic field directions at which the expected Zeeman splitting is observed. While maintaining a fixed magnetic field magnitude, ESR spectra was measured at discrete steps in magnetic field directions. The magnetic field magnitude was chosen to be weak enough that our coils could produce it in any direction, although due to the symmetry of the NV centers, sweeping the magnetic field only along one hemisphere was sufficient. At each field orientation, a microwave drive was swept in frequency, and dips in the florescence indicated resonance with a spin transition. 


We monitored the ESR dip depth and location while sweeping a fixed magnitude of $B_\mathit{ext}$ around in polar [0-90$^\circ$]($\phi$) and azimuthal [0-360$^\circ$]($\Theta$) angles, where the polar axis is defined to be normal to the substrate. The results for a specific diamond nanocrystal, which was $\approx$ 10~$\mu$m away from the edge of the MW loop, are shown in \figref{fig:BField}. Similar behavior was observed in other diamond nanocrystals within 20-30~$\mu$m of the MW loop. \figref{fig:20mTDepth} and \figref{fig:20mTLocation} show the behavior of the ESR dip depth and dip location respectively, while sweeping $B_\mathit{ext}$ in a hemispherical manner with 20~mT magnitude. Such plots help us broadly recognize the regions where the ESR dip depth is a maximum and dip location is a minimum, as shown by the circled regions in these plots. These regions give the approximate orientations of the NV centers in this diamond nanocrystal, where we expect them to behave as $\theta^{2}$ sensors with very high sensitivity when the magnetic field magnitude is increased. Due to the experimentally accessible magnetic field region being close to the pole of the hemispheres, we chose the region nearest the pole for further exploration. \figref{fig:20mTDip} shows the ESR dip in this region which fits well to a Lorentzian lineshape with frequency at $\approx$~2.3~GHz, as expected. 

We further narrowed down the spread of the NV center response in both $\phi$ and $\Theta$ by probing with higher magnetic fields. \figref{fig:80mTLocation} shows such a plot of the behavior of the ESR dip location while sweeping an 80~mT $B_\mathit{ext}$. As can be seen in this plot, the spread of both polar and azimuthal angles narrows down as the external magnetic field is increased. In this case, the spread in the polar angle is much finer ($<$5$^\circ$) than the spread in the azimuthal angle ($<$25$^\circ$), because the field direction is near the pole. The shifted ESR dip at 80~mT at 13.4$^\circ$~$\phi$ and 267$^\circ$~$\Theta$ is shown in \figref{fig:80mTDip}, fitted to a Lorentzian lineshape with the dip frequency at $\approx$~0.63~GHz. 

\begin{figure*}
	\begin{subfigure}{0.49\linewidth}
		\includegraphics[scale=0.32]{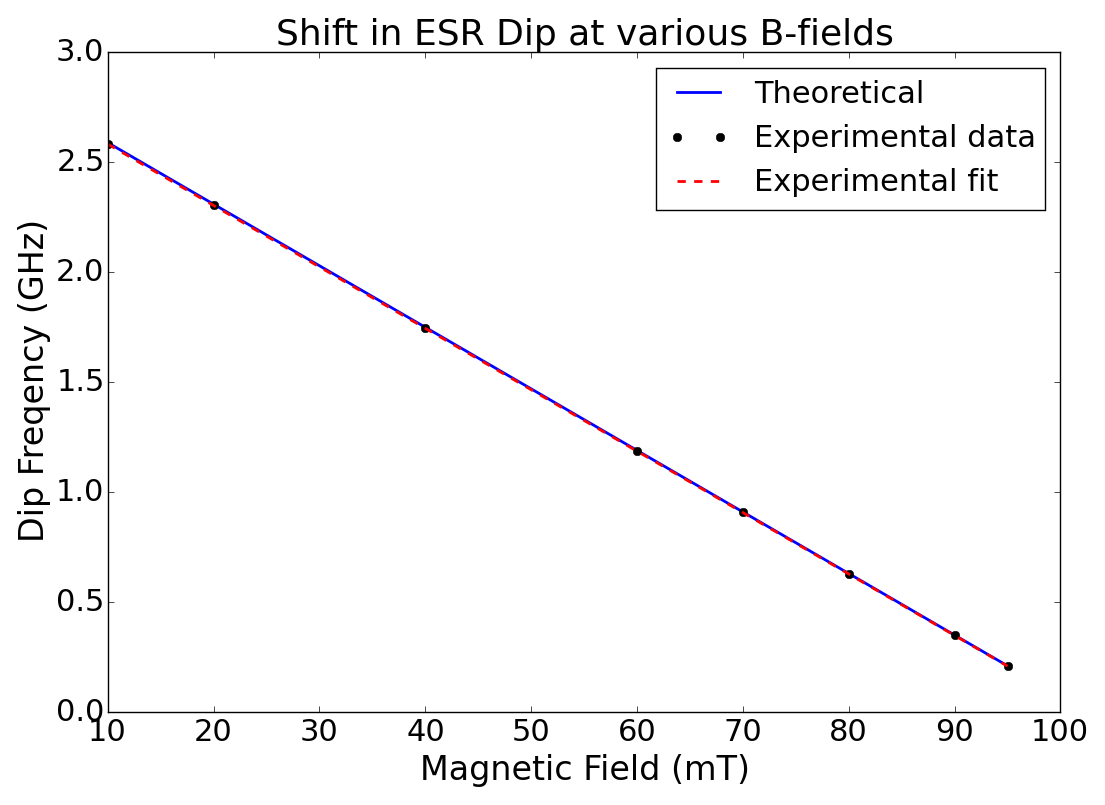}
		\caption{\label{fig:DipShift}}
	\end{subfigure}
	\begin{subfigure}{0.49\linewidth}
		\includegraphics[scale=0.32]{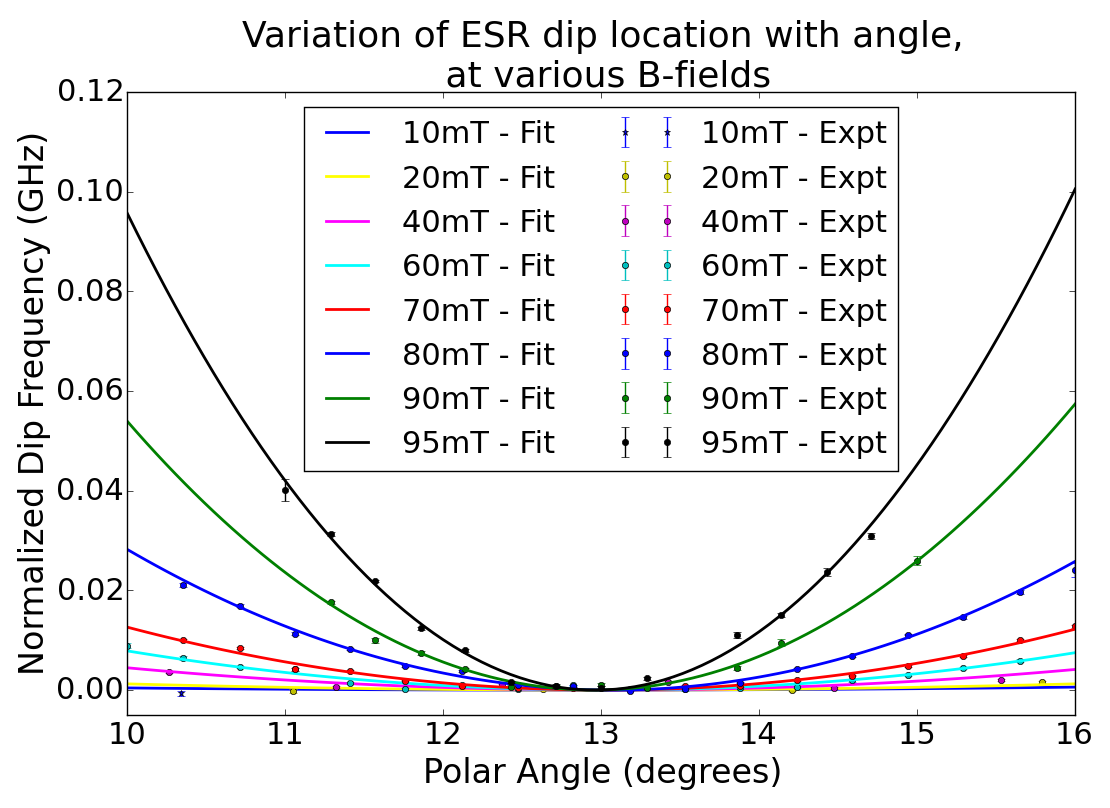}
		\caption{\label{fig:QuadSens}}
	\end{subfigure}
	\begin{subfigure}{0.49\linewidth}
		\includegraphics[scale=0.32]{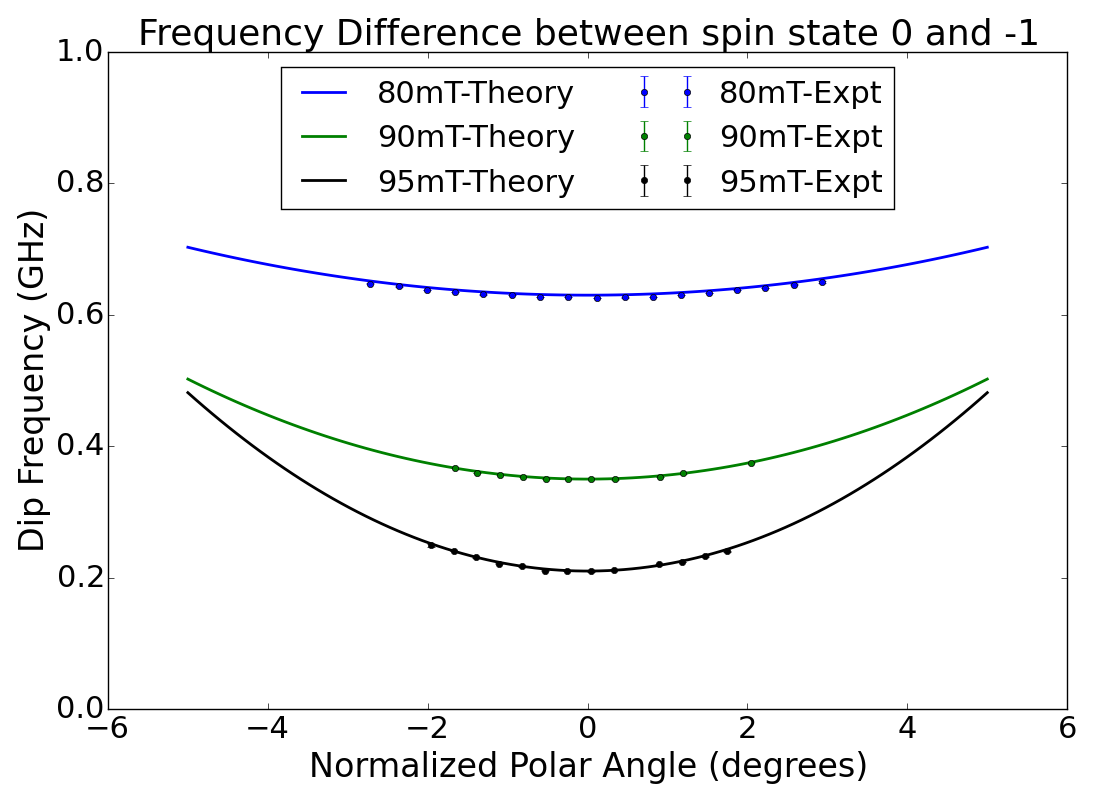}
		\caption{\label{fig:EigDiff-E+T}}
	\end{subfigure}
	\begin{subfigure}{0.49\linewidth}
		\includegraphics[scale=0.32]{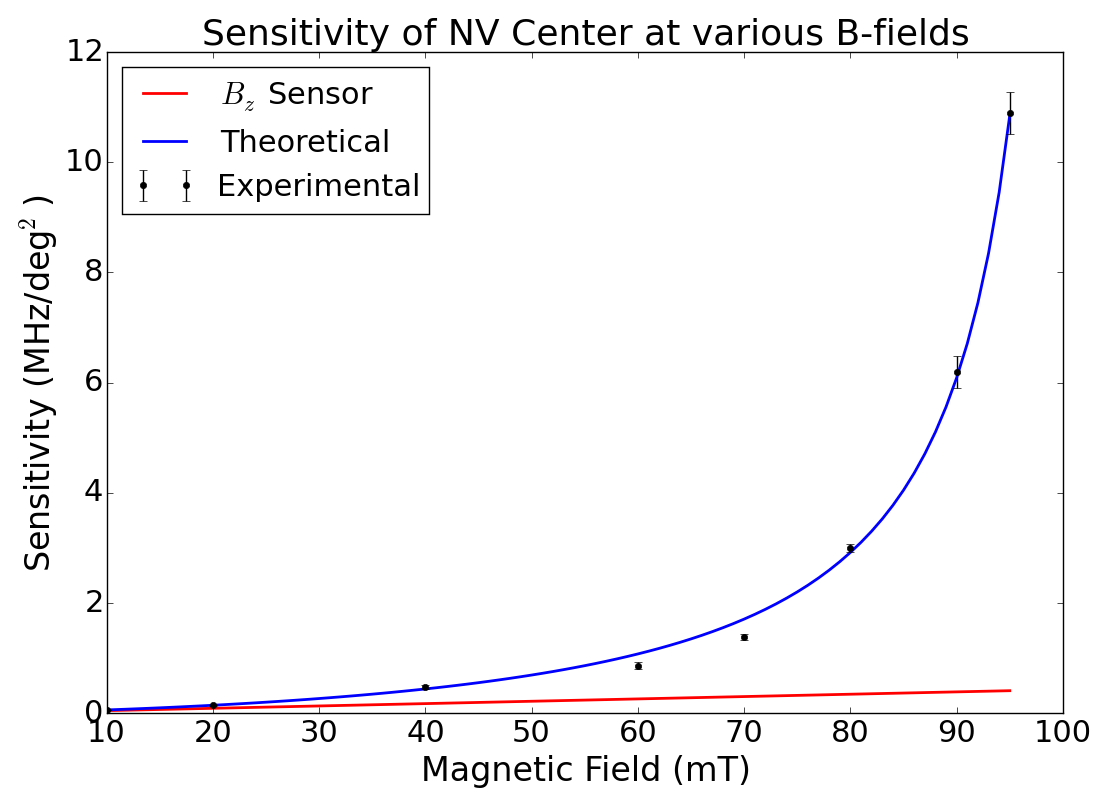}
		\caption{\label{fig:Sens}}
	\end{subfigure}
\caption{Various trends seen with the shift of ESR dip on applying external magnetic field. As predicted (\subref{fig:DipShift}) the ESR dip shifts down in frequency linearly with the $B_\mathit{ext}$ magnetic field and (\subref{fig:QuadSens}) the ESR dip depth and location have a quadratic behavior with angle; dip location is plotted here with respect to the polar angle at various magnetic fields. (\subref{fig:EigDiff-E+T}) The predicted and measured absolute transition frequencies are in excellent agreement. (\subref{fig:Sens}) Plotting the quadratic sensitivity at different magnetic fields shows that the NV center is a much more sensitive $\theta^{2}$ sensor than any `B$_{z}$ sensor', which measures the component of the magnetic field along its axis.}
\label{fig:trends}
\end{figure*}

By sweeping $B_\mathit{ext}$ between 0-95~mT in $\Theta$ and $\phi$ space, the ESR dip was observed to shift down linearly in frequency on increasing the magnitude of $B_\mathit{ext}$, as seen in \figref{fig:DipShift}. Fitting this experimental data to a straight line, we get an $x$-intercept of 102.46~mT, which is very close to the expected value of $B_\mathit{zfs}$ = 102.5~mT, and a $y$-intercept of 2.863~GHz, which is very close to zero-field splitting, $\Delta$ = 2.87~GHz. At several magnitudes of $B_\mathit{ext}$, the variation of the ESR dip frequency, with respect to $\phi$ and $\Theta$, was found to be quadratic, as in \figref{fig:QuadSens} and \figref{fig:EigDiff-E+T}. The coefficients of the $\theta^2$ dependence at different magnitudes of $B_\mathit{ext}$ is shown as the sensitivity in \figref{fig:Sens} and are in excellent agreement with the expected theoretical behavior. This experimentally verifies that the NV centers are highly sensitive $\theta^2$ sensors, with sensitivity that increases as $B_\mathit{ext}$ increases, and diverges as it approaches $B_\mathit{zfs}$. For comparison, if the NV center were only affected by the component of the magnetic field along its $\hat{z}$ axis, the $\theta^2$ sensitivity would be linear with $B_\mathit{ext}$, as in~\figref{fig:Sens}.

\section{\label{sec:Conclusion}Conclusion}

In conclusion, we have shown that NV centers in a diamond nanocrystal can be used as a high-sensitivity $\theta^2$ sensors. To use a particular nanocrystal for this purpose, a weak magnetic field is first swept over a wide range of angles to identify the orientations of the NV centers. After increasing the magnitude of the field and aligning it near the NV center $\hat{z}$ axis, the sensitivity of the NV center ESR frequency to $\theta^2$ is greatly enhanced, even diverging as $B_\mathit{ext}$ approaches $B_\mathit{zfs} \approx 102.5$~mT, as predicted in our previous work~\cite{Durso11}. The ability of the NV centers to be highly sensitive $\theta^2$ sensors is critical for their use in making QND measurements of e.g. the number state of a torsional harmonic oscillator. This class of measurements may be valuable in a variety of applications such as gravity wave detection and probing quantum behavior of NMOs.

\bibliography{nmo}

\end{document}